\documentclass{ifacconf}

\usepackage{algorithm}

\usepackage{algorithmic}
\usepackage{comment}
\usepackage{xcolor}
\usepackage{makecell}
\usepackage{amsmath}
\usepackage{amssymb}
\usepackage{mathtools}
\usepackage{graphicx}      
\usepackage{subcaption}
\usepackage{threeparttable}
\usepackage[T1]{fontenc}
\newtheorem{remark}{Remark}
\newtheorem{definition}{Definition}
\usepackage{natbib}        
\begin{document}
\begin{frontmatter}

\title{Adaptive Uncertainty Quantification for Scenario-based Control Using Meta-learning of Bayesian Neural Networks\thanksref{footnoteinfo}} 

\thanks[footnoteinfo]{This work was financially supported by the United States National Science Foundation under award CMMI-2302219.}

\author[First]{Yajie Bao} 
\author[Second]{Javad Mohammadpour Velni} 

\address[First]{Intelligent Fusion Technology, Inc., 
   Germantown, MD 20874 USA (e-mail: yajie.bao@intfusiontech.com).}
\address[Second]{Clemson University, 
   Clemson, SC 29634 USA (e-mail: javadm@clemson.edu)}

\begin{abstract}                
Scenario-based optimization and control has proven to be an efficient approach to account for system uncertainty. In particular, the performance of scenario-based model predictive control (MPC) schemes depends on the accuracy of uncertainty quantification. However, current learning- and scenario-based MPC (sMPC) approaches employ a single time-invariant probabilistic model (learned offline), which may not accurately describe time-varying uncertainties. Instead, this paper presents a model-agnostic meta-learning (MAML) of Bayesian neural networks (BNN) for adaptive uncertainty quantification that would be subsequently used for adaptive-scenario-tree model predictive control design of nonlinear systems with unknown dynamics to enhance control performance. In particular, the proposed approach learns both a global BNN model and an updating law to refine the BNN model. At each time step, the updating law transforms the global BNN model into more precise local BNN models in real time. The adapted local model is then used to generate scenarios for sMPC design at each time step. A probabilistic safety certificate is incorporated in the scenario generation to ensure that the trajectories of the generated scenarios contain the real trajectory of the system and that all the scenarios adhere to the constraints with a high probability. 
Experiments using closed-loop simulations of a numerical example demonstrate that the proposed approach can improve the performance of scenario-based MPC compared to using only one BNN model learned offline for all time steps.
\end{abstract}


\end{frontmatter}
\vspace{-2mm}
\section{Introduction}
\vspace{-2mm}
Uncertainty is ubiquitous in control systems and may lead to system constraint violation and/or performance deterioration of a designed controller. Scenario-based optimization, especially scenario-based model predictive control (MPC), has been developed to account for system uncertainties by representing uncertainties with a \textit{scenario tree}, and reduce the conservativeness inherent to open-loop robust MPC by introducing recourse into the optimal control problem \citep{bemporad2002model}. However, generating scenarios offline based on worst-case uncertainty descriptions obtained \textit{a priori} can limit the performance of sMPC. To generate a scenario tree that can accurately represent the evolution of uncertainties depends on an adequate description of uncertainties, which is generally not available in practice. Using data-driven models of uncertain nonlinear systems for MPC design (aka learning-based MPC) has attracted increasing attention \citep{hewing2020learning,mesbah2022}, as machine learning has proven to be effective for modeling time-varying and/or hard-to-model dynamics. However, there are still challenges for data-driven modeling to provide accurate and generalizable models which are critical for sMPC performance. One particular challenge is how to cope with the discrepancy between the application environment and the data-collecting environment (aka data drift \citep{https://doi.org/10.48550/arxiv.2012.09258}), which can degrade model accuracy and subsequently control performance. Therefore, adapting data-driven models to new environments in real-time is necessary to ensure a desired control performance \citep{bao2020online,bao2021physics}.


Gaussian process regression (GPR) \citep{rasmussen2003gaussian} is the most widely-used approach for data-driven characterization of model uncertainty in learning-based MPC \citep{koller2018learning,hewing2019cautious,soloperto2018learning, bonzanini2021fast}. While experiments demonstrated that GPR was able to capture structural model uncertainty, GPR suffers from a cubic complexity to data size, which may restrict the size of data used for efficacy offline training and online evaluation. Moreover, GPR assumes that the model uncertainty can be described by a joint Gaussian distribution, which may be invalid for applications. Alternatively, Bayesian neural networks (BNNs) have been increasingly used to quantify uncertainties \citep{bao2020cdc,bao2022safe} and learning-based MPC \citep{BAO2022RNC,bao2022learning}. BNNs treat the weights of deterministic neural network (NN) models as random variables and provide estimates of the posterior distributions conditioned on a dataset. Compared to GPR, 
BNNs can model both epistemic and aleatoric uncertainties with arbitrary distributions, be trained efficiently using `Bayes by Backprop' \citep{blundell2015weight}, and be quickly evaluated without using the training dataset to compute kernel matrices. BNNs can be viewed as an ensemble of deterministic networks (ANN models) combined by the posterior distributions and can provide robust predictions using cheap model averaging \citep{carbone2020robustness}. However, BNNs suffer from high computational cost and noisy gradient as a result of estimating the evidence lower bound (ELBO) from a single sample of weights \citep{jospin2020hands}, which increases the difficulty in adapting BNN models on a small batch of data in real time.

To tackle the challenges of online adaptation of BNNs, we resort to \textit{model-agnostic meta-learning (MAML)} methods. MAML \citep{finn2017model} is compatible with BNNs that are trained with gradient descent (GD). Moreover, MAML trains models that are easy to be fine-tuned, and enable fast adaptation of deep networks with good generalization performance. Specifically, MAML explicitly trains a model on various tasks such that using a small amount of training data to update the model parameters for a small number of GD steps produces a good model for a new task. In the case of control design, predicting system outputs/behaviors at each time step is viewed as a different task such that the sudden changes (at any time step) in the environments can be coped with through model adaptation. Moreover, different from \cite{bao2020online} that requires a batch of closed-loop data for online transfer learning and \cite{bao2021physics} that uses adaptive sliding mode control to derive the updating law, the adaptation is performed by a parameterized updating law which takes recent system trajectory and the global model parameters as inputs and outputs the adapted parameters of the precise local model. 

\begin{figure}
    \centering
    \includegraphics[width=\columnwidth]{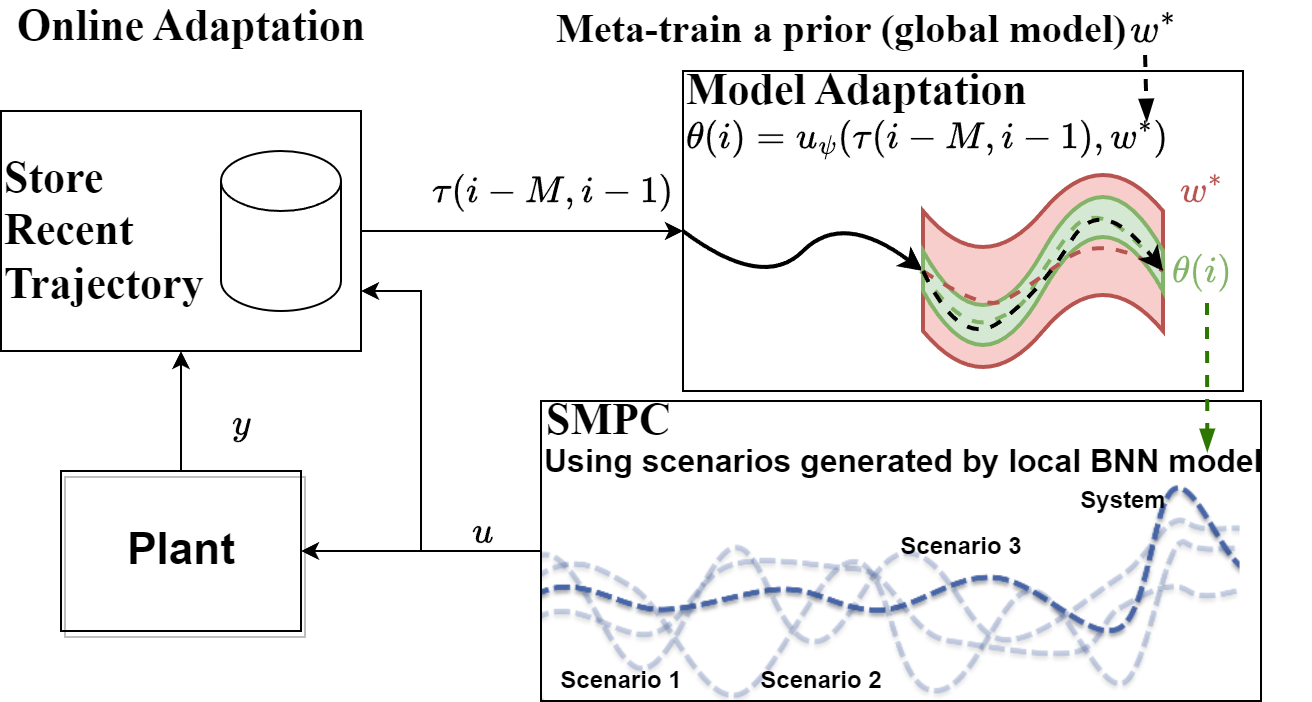}
    \caption{Schematic of adaptive sMPC using MAML of BNNs, where a global BNN model is adapted online using recent trajectories for scenario generation.}
    \label{fig:maml_closed}
    \vspace{-2mm}
\end{figure}

For sMPC design using BNN models, a BNN model is initially learned to model state- and input-dependent uncertainties, and the statistics of the BNN predictions are then used to generate adaptive scenarios online. Additionally, the sMPC \citep{BAO2022RNC} using BNNs improved the robust control performance with respect to sMPC with a fixed scenario tree and with respect to an adaptive scenario-based MPC using Gaussian process regression, by realizing a less conservative estimation of the model uncertainty. This paper aims to further improve the control performance by adapting the BNN model online. Fig. \ref{fig:maml_closed} illustrates the proposed scheme for adaptive sMPC design using model-agnostic meta-learning of BNNs.

Additionally, \cite{finn2018learning} proposed a meta-learning approach for adaptive control using reinforcement learning with deterministic models. To quantify uncertainty, \cite{harrison2018control} used meta-learning of Bayesian linear regression in the feature space to model long-term dynamics and employed an LQR-type control scheme with a linearized dynamics model for uncertainty-aware control. Instead, \cite{richards2021adaptive} proposed adaptive control-oriented meta-learning to directly train a parametric adaptive controller using closed-loop simulations, which adapts well to each model of an ensemble constructed from past inputs/outputs data. Different from the aforementioned works, this paper presents a MAML approach for fast online adaptation of BNN-based uncertainty model and then employs the adapted model for sMPC design of uncertain constrained nonlinear systems. \textit{The main contribution of this paper lies in presenting MAML of BNN models for time-varying, state- and input-dependent uncertainty quantification.} 

\vspace{-2mm}
\section{Problem Formulation and Preliminaries}
\vspace{-2mm}
Consider a constrained, discrete-time nonlinear system with state- and input-dependent uncertainty of the form 
\begin{subequations}
\label{eq:sys}
\begin{align}
    x(k+1) &= f(x(k),u(k)) + g(x(k),u(k)), \label{eq:mdl}\\
    x&\in\mathcal{X}, u\in \mathcal{U}, \label{eq:cons}
\end{align}
\end{subequations}
where $x$ is the state, $u$ is the control input, and $k\in\mathbb{N}$ is the time instant; $f:\mathcal{X}\times \mathcal{U}\rightarrow\mathcal{X}$ describes a known, Lipschitz continuous model \eqref{eq:mdl} while $g:\mathcal{X}\times \mathcal{U}\rightarrow\Omega$ represents \textit{a priori} unknown model error term which is assumed to be Lipschitz continuous; $\mathcal{X}\subseteq\mathbb{R}^{n_{x}}$ and $\mathcal{U}\subseteq\mathbb{R}^{n_{u}}$ in \eqref{eq:cons} are the constraint sets of the states and inputs, respectively. $\mathcal{X}$ is assumed to be convex. The initial state $x(0)=x_{0}$.

\begin{remark}
The known model $f$ can be a first principles-based model or obtained as a data-driven model. One interesting class of models is linear parameter-varying (LPV) models. LPV models use a linear structure to capture time-varying and nonlinear dynamics, which allows the development of computationally efficient control design methods. Moreover, data-driven methods have been increasingly developed for the global identification of state-space LPV models (e.g., see \cite{rizvi2018state,bao2022overview}).
\end{remark}

Assuming a dataset $\mathcal{D}=\{(x^{(i)},u^{(i)},x^{(i+1)})|i=1,\cdots,N_{\mathcal{D}}\}$ that covers the entire feasible space $\mathcal{X}\times \mathcal{U} \times \mathcal{X}$ has been collected from the real system \eqref{eq:sys}, this paper aims to learn an adaptable BNN model of $g$ using MAML for sMPC design with safety guarantees. 

The closed-loop system model using sMPC can be expressed by
\begin{equation}
\label{eq:sys_closed}
\begin{split}
x(k+1) &= f(x(k),\kappa(x(k)))+g\left(x(k),\kappa(x(k))\right) \\
&\triangleq\Phi_{\kappa}(x(k))
\end{split}
\end{equation}
with $\kappa:\mathcal{X}\times\mathbb{N} \rightarrow\mathcal{U}$ denoting the control law. We use $\mathbf{x}(k|x_{0})\triangleq \{x(1),\cdots,x(k) \}$ to denote the solutions to the model \eqref{eq:sys_closed} given the initial state $x_{0}$.

\begin{definition}[\cite{koller2018learning}]\label{def:safe}
Given $x_{0}\in \mathcal{X}$, the system (\ref{eq:mdl}) is said to be \textbf{safe} under a control law $\kappa$ if
\begin{equation}
\label{cond:safe}
\forall k \in \mathbb{N}, \Phi_{\kappa}(x(k))\in\mathcal{X}, \kappa(x(k))\in \mathcal{U}.
\end{equation}

The system (\ref{eq:mdl}) is \textbf{$\delta$-safe} under $\kappa$ if
\begin{equation}
\label{cond:delta_safe}
\forall k \in \mathbb{N}, \Pr\left[\Phi_{\kappa}(x(k))\in\mathcal{X}, \kappa(x(k))\in \mathcal{U}\right]\geq \delta,
\end{equation}
where $\Pr[\cdot]$ denotes the probability of an event.
\end{definition}

Next, we introduce the preliminaries of meta-learning and BNNs for uncertainty quantification.

\subsection{Meta-learning}

Meta-learning seeks the adaptation of machine learning models to unseen tasks that are vastly different from trained tasks \citep{peng2020comprehensive}. Specifically, given a distribution  $p(\mathcal{T})$ of tasks, meta-learning aims to solve $\min_{w}\mathbb{E}_{\mathcal{T}\sim p(w)}\mathcal{L}(\mathcal{D};w)$ where $\mathcal{T}=\{\mathcal{D},\mathcal{L}\}$ denotes a task consisting of a dataset $\mathcal{D}$ and a loss function $\mathcal{L}$, and $w$ denotes the meta-knowledge (such as the choice of optimizer and the function class for a task). 

Meta-learning consists of an inner level for base learning and an outer level as the meta-learner. At the inner level, a new task with a dataset $\mathcal{D}_{source}^{(i)}$ from a set of $M$ source tasks $\mathcal{D}_{source}=\{(\mathcal{D}_{source}^{train},\mathcal{D}_{source}^{val})^{(i)}\}_{i=1}^{M}$ is presented, and the agent aims at quickly learning the associated concepts with the task from the training observations, i.e., finding $\rho^{*}=\arg\max_{\rho}\text{log}~p(\rho|w^{*},\mathcal{D}_{source}^{train})$. This quick adaptation is facilitated by knowledge accumulated across earlier tasks \citep{huisman2021survey} (aka, meta-knowledge). At the outer level, the learner updates the inner level algorithm such that an outer objective function (e.g., generalization performance) is improved for learning meta-knowledge, i.e., $w^{*}=\arg\max_{w}\textup{log}~p(w|\mathcal{D}_{source})$. In this way, meta-learning can continually perform self-improvement as the number of tasks increases. 

To quantify uncertainties, we use BNNs as the base learner for MAML. For fast online adaptation, rather than collecting a batch of closed-loop data for online transfer learning \citep{bao2020online}, we adopt an unsupervised domain adaptation approach, and the online adaptation will be based on the updating law, which avoids noisy gradients of training BNNs by backpropagation. The details of MAML of BNNs will be provided in Section \ref{sec:meta}.

\subsection{Bayesian Neural Networks}
\label{sec:bnn_data}
The key component of a BNN is the DenseVariational layer which approximates the posterior density $p(W|\mathcal{D})$ of the parameters $W$ by variational inference (VI) given a prior density $p(W)$ where $\mathcal{D}$ denotes a dataset. A reparameterization trick is employed to parameterize $q(W_j;\theta_{j})$ with parameters $\theta_{j}$ for approximating $p(W|\mathcal{D})$,  i.e., 
\begin{equation}
\label{eq:repara}
W_j=\mu_{W{_j}} + \sigma_{W_{j}} \odot \epsilon _{W_j}
\end{equation}
where $\theta_{j} = \{\mu_{W_{j}}, \sigma_{W_{j}}\}$ in this case; $\odot$ denotes element-wise multiplication; $\epsilon _{W_j} \sim \mathcal{N}(0, I)$. A multi-layer, fully connected BNN is used to model the unknown vector-valued function $g$. $\mathcal{D}_{g}=\{\mathrm{x}^{(i)} = (x^{(i)},u^{(i)}),g^{(i)}|i=1,\cdots,N_{g}\}$ for training the BNN is obtained by computing $g^{(i)} =x^{(i+1)}-f(x^{(i)},u^{(i)})=g(x^{(i)},u^{(i)})$ on the dataset $\mathcal{D}$, where $\mathrm{x}^{(i)}$ denotes the input to the BNN, and $g^{(i)}$ is the uncertainty to be predicted by the BNN. The details of training BNNs can be found in \cite{bao2020cdc}. With the trained BNN, the probability density function of $\hat{g}$ for a given $(x(k), u(k))$ can be approximated by sampling from the posterior distributions of weights using Monte Carlo (MC) methods and computing $\hat{g}$ for each set of sampled weights. 
\subsection{Scenario-based MPC Design Approach} 
At the time instant $k$, the stochastic MPC minimizes 
\begin{align}
\label{eq:stochastic-mpc}
    \mathbb{E}\left\{\sum_{i=0}^{N-1}\ell(x(i|k),u(i|k))+V_{N}(x(N|k))\right\} 
\end{align}
where $\mathbb{E}\{\cdot\}$ is the expectation operator over the random vector sequence $\mathbf{g}=\{g(0), \cdots, g(N-1)\}$. The uncertainties of $g$ are propagated forward through the prediction model \eqref{eq:mdl}, making it difficult to derive the closed-form probability density function of $\mathbf{g}$. To evaluate the cost in \eqref{eq:stochastic-mpc}, scenario-based MPC (sMPC) uses a tree of discrete scenarios to represent the uncertainty evolution of a system. Consequently, the scenario-based optimal control problem for an uncertain system at time step $k$ can be formulated as follows:
\begin{subequations} \label{eq:ocp-general}
\begin{align}
    \min_{x^{j},u^{j}} &~~ \sum_{j=1}^{S} p^{j} \left[ \sum_{i=0}^{N-1} \ell\left(x^{j}(i|k),u^{j}(i|k)\right) + V_N\left(x^{j}(N|k)\right) \right] \label{eq:mpc-cost}\\
    \text{s.t.} &~~ x^{j}(i+1|k) = f\left(x^{j}(i|k),u^{j}(i|k)\right) + \hat{g}^{j}(i|k), \label{eq:dynamics}\\
    &~~ \left(x^{j}(i|k),u^{j}(i|k)\right) \in \mathcal{X} \times \mathcal{U}, \label{cons:smpc}\\
    &~~ x^{j}(0|k) = x(k), \\
    &~~ u^{j}(i|k) = u^{l}(i|k) ~ \text{if} ~ x^{p(j)}(i|k) = x^{p(l)}(i|k), \label{eq:non-anticipativity}
\end{align}
\end{subequations}
where the superscript $j \in \{1,\ldots,S\}$ is the index of the scenario; $p^{j}$ is the probability of the $j$-th scenario; $\ell\left(x^{j}(i|k),u^{j}(i|k)\right)$ and $V_N\left(x^{j}(N|k)\right)$ are the stage cost and terminal cost for the $j$-th scenario, respectively; $N$ is the prediction horizon length; $\hat{g}^{j}$ is the uncertainty realization in the $j$-th scenario; and \eqref{eq:non-anticipativity} is the \emph{non-anticipativity} constraint. The control law can be determined by the solution to \eqref{eq:ocp-general} as
    $\kappa\left(x(k)\right) = \mathbf{u}^\star(0|k)$.

\vspace{-2mm}
\section{Learning-based Scenario Generation for sMPC Design}
\label{sec:scenario_generate}
\vspace{-2mm}
At each time step $k$, we draw $\bar{N}_{\text{MC}}$ samples from normal distributions and calculate weights $W^{(i)}$ by the transformation \eqref{eq:repara} to the $i$-th sample. Although Lemma 1 in \cite{10146017} states that the trajectories of the $\bar{N}_{\text{MC}}$ sampled models encompass the system trajectory, $\bar{N}_{\text{MC}}$ can be too large to be practical for online optimization. To reduce the number of scenarios, we instead evaluate the estimate $\hat{g}^{(i)}(k)$ using $W^{(i)}$, then calculate the sample mean $\hat{\mu}_{g(k)} =\frac{1}{\bar{N}_{\text{MC}}} \sum_{i=1}^{\bar{N}_{\text{MC}}}\hat{g}^{(i)}(k)$ and standard deviation $\hat{\sigma}_{g(k)} = \sqrt{\frac{1}{\bar{N}_{\text{MC}}} \sum_{i=1}^{\bar{N}_{\text{MC}}}(\hat{g}^{(i)}(k)- \hat{\mu}_{g(k)})^{\top}(\hat{g}^{(i)}(k)- \hat{\mu}_{g(k)})}$, and use $\hat{\mu}_{g(k)}$, $\hat{\mu}_{g(k)}+m^{j}\hat{\sigma}_{g(k)}$, $\hat{\mu}_{g(k)}-m^{j}\hat{\sigma}_{g(k)},j=1, \cdots, \frac{S-1}{2}$ where $m^{j}$ are the tuning parameters and $S$ is the number of scenarios. The probabilities of the scenarios are calculated using the moment matching method \citep{hoyland2001generating} to maintain the original statistical properties. To save computational cost, we only update the uncertainty estimation when solving \eqref{eq:ocp-general} and fix the scenarios over the prediction horizon. Specifically, we use the solution $u^{*}(1|k-1)$ to \eqref{eq:ocp-general} at $k-1$ and the state $x(k)$ to estimate uncertainty $\hat{g}(k)$, and $\hat{g}(i|k)=\hat{g}(k), i=0,\cdots,N-1$ for \eqref{eq:ocp-general} at $k$. This approach is more tractable than considering time-varying uncertainties and adaptive scenarios within the prediction horizon, as the uncertainties are input-dependent and the control input sequence in the prediction horizon are decision variables of the sMPC problem. When the uncertainties do not change significantly within the prediction horizon, fixing the uncertainty estimation is reasonable and less conservative than using worst-case error bounds.


\section{Meta-learning of BNN for sMPC Design}\label{sec:meta}
\begin{figure}[!htbp]
     \centering
     \includegraphics[width=1.3\columnwidth]{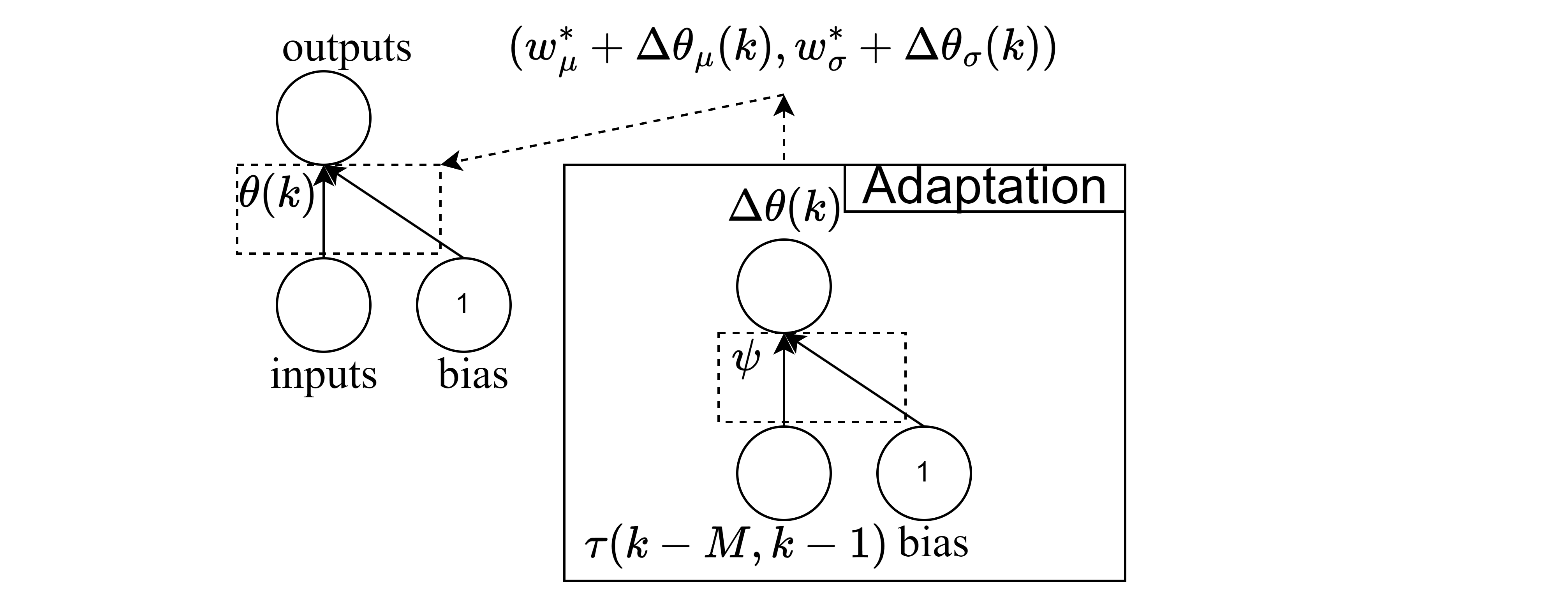}
     \caption{Online adaptation of the BNN model.}
     \label{fig:online-adapt}
     \vspace{-2mm}
 \end{figure}

\vspace{-2mm}
In this section, we present the model-agnostic meta-learning (MAML) approach for fast adaptation of BNN models. In particular, MAML learns a parameterized adaptation law that transforms a global BNN model (i.e., the meta-knowledge on a distribution of tasks) into a local BNN model with improved accuracy at each time step, to provide a tighter uncertainty quantification for scenario generation and thus improves the control performance. Specifically, meta-learning aims to find meta-knowledge that is useful for a distribution of tasks $p(\mathcal{T})$, i.e.,  
\begin{equation}
    w^{*} = \arg\min_{w}\mathbb{E}_{\mathcal{T}\sim p(\mathcal{T})}\mathcal{L}(\mathcal{D};w)
\end{equation}
where $\mathcal{D}$ denotes a dataset, $\mathcal{L}$ is the loss function, and $\mathcal{T}=\{\mathcal{D},\mathcal{L}\}$. To adapt to a specific task $\mathcal{T}^{(i)}$, meta-learning finds task-specific parameters $\theta^{(i)}$ by solving
\begin{equation}
    \min_{\theta^{(i)}} \mathcal{L}^{(i)}(\mathcal{D}^{(i)},\theta^{(i)};w^{*})
\end{equation} based on the meta-knowledge, which facilitates quick adaptation. In our case, inspired by \cite{finn2018learning}, we view the model adaptation at each time step as a different task and obtain the parameters of the model at time step $k$ by $\theta(k)=w^{*}+\Delta\theta_{\psi}(\tau(k-M,k-1))$ where $\tau(k-M,k-1)=\left(x(k-M),u(k-M),\cdots,x(k-1),u(k-1)\right)$ denotes the trajectory of the system in the last $M$ time steps and $\Delta\theta_{\psi}$ is the step size function represented by an ANN with parameters $\psi$. Fig. \ref{fig:online-adapt} demonstrates the online adaptation of the global BNN model parameterized by $w^{*}$ to the local BNN model parameterized by $\theta(k)$. Then, the meta-learning problem is 
\begin{align}
\min_{w,\psi} ~ &\mathbb{E}_{i\sim \mathcal{U}\{M,\cdots,N-K\}} \left[\mathcal{L}(\tau(i,i+K),\theta(i))\right] \\
\textup{s.t.} ~ & \theta(i)=w+\Delta\theta_{\psi}(\tau(i-M,i-1)).
\end{align}
Moreover, the loss function of a task is
\begin{align*}
    \mathcal{L}&=\mathcal{L}(\tau(i,i+K),\theta(i)) \\
    &= \frac{1}{K}\sum_{k=1}^{K} \mathcal{L}(\hat{x}(i+k), x(i+k);\theta(i+k-1)).
\end{align*}
It is noted that the loss function is evaluated on the future transitions such that the prediction errors of the adapted models are minimized in the next $K$ time steps. The training process is summarized in Algorithm \ref{alg:meta}, and the testing process is presented in Algorithm \ref{alg:ctrl}.
\begin{algorithm}[!htbp]
 \caption{\label{alg:meta}Learning the meta-knowledge} 
 \begin{algorithmic}[1]
 \REQUIRE learning rate $\alpha,\beta\in\mathbb{R}^{+}$ for $\psi,w$, respectively; number of tasks $N_{\mathcal{T}}$; dataset $\mathcal{D}$
  \STATE Initialize $w$
  \WHILE{not done}
  \FOR {$i = 1$ to $N_{\mathcal{T}}$}
  \STATE Sample $i\sim \mathcal{U}\{M,\cdots,N-K\}$ and have $\tau(i-M,i-1),\tau(i,i+K)$ 
  \STATE $\theta(i)=w +\Delta\theta_{\psi}(\tau(i-M,i-1))$
  \STATE $\mathcal{L}_{j}=\mathcal{L}(\tau(i,i+K),\theta(i))$
  \STATE $\psi\xleftarrow[]{}\psi-\alpha \bigtriangledown _{\psi}\mathcal{L}_{j}$ 
  \ENDFOR
  \STATE meta-update: $w\xleftarrow[]{}w-\beta \frac{1}{N_{\mathcal{T}}} \sum_{j=1}^{N_{\mathcal{T}}} \bigtriangledown _{w}\mathcal{L}_{j}$ 
  \ENDWHILE \\
 \RETURN $w^{*}$ and $\Delta\theta_{\psi}$ 
 \end{algorithmic} 
 \end{algorithm}
 
 \begin{algorithm}[!htbp]
 \caption{\label{alg:ctrl}Online adaptive control design approach}
 \begin{algorithmic}[1]
 \REQUIRE meta-knowledge $w^{*}$; update rule $\Delta\theta_{\psi}$; experience $\tau(-M,-1)$ and time steps $\bar{N}$ for control
  \FOR {$k = 0$ to $\bar{N}$}
  \STATE Adapt model by $\theta(k)=w^{*} +\Delta\theta_{\psi}(\tau(k-M,k-1))$
  \STATE Compute control input $u(k)$ using the model with $\theta(k)$ 
  \STATE Apply $u(k)$ and collect data $(x(k),u(k),x(k+1))$ for validating/fine-tuning the model
  \ENDFOR
 \end{algorithmic} 
 \end{algorithm}

We directly learn a set of parameters of model weights posteriors as the meta-knowledge and adapt to task-specific model weights posteriors by modeling the task-specific parameters of the model weights posteriors as a function of the past trajectory and the meta-knowledge for fast adaptation. Moreover, we initialize the meta-knowledge with the model learned in Section \ref{sec:bnn_data} to facilitate learning.

Prior distributions affect the accuracy of the BNN models, and their selection has been studied by \cite{fortuin2021bayesian}. A proper prior is usually unknown or hard to choose. For MAML of BNNs, we directly use the learned posterior as the prior of the BNN weights. That way, the local model adjusts the posteriors in a way that is as close as possible to the posteriors of the global model, i.e.,
\begin{equation}
\begin{split} 
\min _{w,\psi} \Big(&\mathbb{E}_{q(W;w,\psi)}\left[\log q(W;w,\psi)\right] -
\mathbb{E}_{q(W;w,\psi)}\left[\log p(W;\theta_{0}) \right]\\& -\mathbb{E}_{q(W;w,\psi)}\left[\log p(\mathcal{D}|W) \right]\Big), 
\end{split}
\label{eq:cost2}
\end{equation}
where $\theta_{0}$ denotes the parameters of the learned posterior. 
\vspace{-2mm}
\section{Experimental Results and Validation}
\vspace{-2mm}
Consider the following nonlinear system
\begin{equation} \label{exp:num}
    \left\{\begin{array}{l}
\dot{x}_{1}=10x_{1}-x_{1}x_{2}^{2}+0.5x_{1}^{2}+0.5x_{1}u_{1}+0.5u_{2}\\ 
\dot{x}_{2}=-x_{2}+0.1x_{1}^{2}+3x_{1}^{2}x_{2}-x_{1}x_{2}u_{1}
\end{array}\right.
\end{equation}
with the state and input constraints 
\begin{align*}
\begin{split}
    &-5\leq x_{1}\leq3,~~~~0\leq x_{2}\leq10; \\
    &-1\leq u_{1}\leq 1,~-1\leq u_{2}\leq 1.
    \end{split}
\end{align*}
The system model \eqref{exp:num} is assumed to be unknown but we can collect data from the system for modeling and control design purposes. 

\subsubsection{Plant-model Mismatch Modeling}
\begin{figure}
    \centering
    \begin{subfigure}{0.49\columnwidth}
    \centering
    \includegraphics[width=\columnwidth]{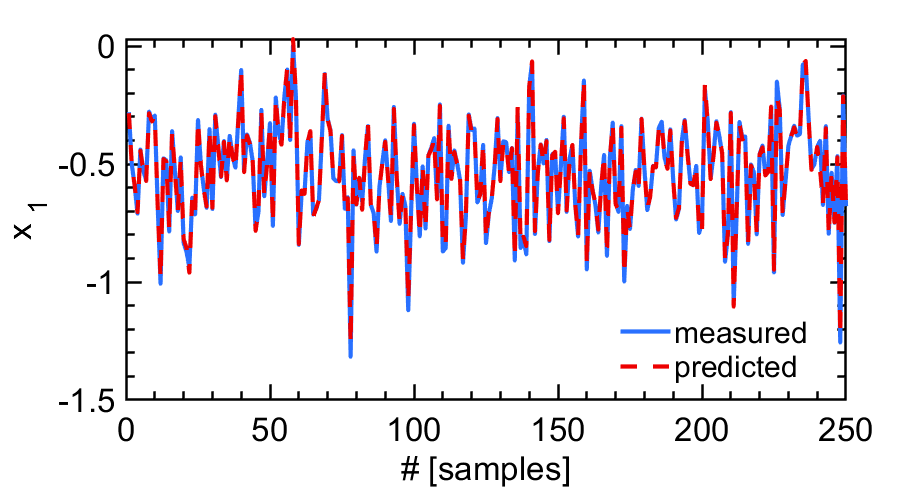}
    \caption{$\textup{BFR}_{x_{1}} = 93.94\%$.}
    \end{subfigure}
    \begin{subfigure}{0.49\columnwidth}
    \centering
    \includegraphics[width=\columnwidth]{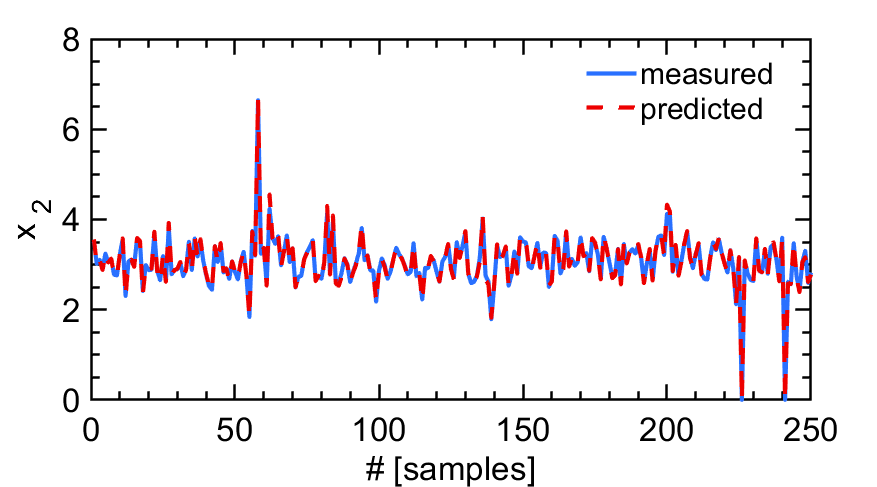}
    \caption{$\textup{BFR}_{x_{2}} = 92.77\%$.}
    \end{subfigure}
    \caption{Validation results of the nominal model for the second example.}     \label{fig:lpv-ann}
    \vspace{-2mm}
\end{figure}

We applied a random input sequence drawn from the uniform distribution $U(-0.5,0.5)$ to the system and collected a dataset $\mathcal{D}=\{u^{(i)},x^{(i)},x^{(i+1)}|i=1,\cdots, 1000\}$ with the sampling time of $0.1$ s. Additionally, the dataset is randomly split into training and testing sets by the ratio of $75\%/25\%$. First, using the approach \citep{bao2020identification}, we learned an ANN-based linear parameter-varying (LPV) model as the nominal model, i.e., $f(x(k),u(k))=A(\rho(k))x(k)+B(\rho(k))u(k)$ where $\rho = [x_{1};x_{1}x_{2}]$ are used as the scheduling variables. In particular, we use two one-layer ANNs without activation functions to represent matrix function $A$ and $B$, respectively. Fig. \ref{fig:lpv-ann} shows the validation results of the ANN-based LPV model. 

Then, we used the proposed MAML of BNN to model the mismatch between the system and the nominal model using the dataset $\mathcal{D}_{g} = \{(x^{(i)},u^{(i)}),x^{(i+1)}-f(x^{(i)},u^{(i)})|i=1,\cdots,1000\}$. 

\textbf{Technical details for MAML of the BNN:} We consider the form $g(x(k),u(k))=h(x(k))[\rho(k);u(k)]$. We used one DenseVariational layer without an activation function fully connected to a three-layer fully-connected ANN with the ELU as activation functions to represent $h$. The prior is $p(W^{4})= \pi\mathcal{N}(0, (\sigma_{1})^{2}) + (1-\pi)\mathcal{N}(0, (\sigma_{2})^{2})$ with $\pi=0.5$, $\sigma_{1}=1.5$, and $\sigma_{2}=0.1$. Each of the two hidden layers has $8$ units. Moreover, we used a fully-connected layer to model the updating rule $\delta \theta (\tau(k-M,k-1))$, and $M=5$. Additionally, we only update the posteriors of the weights and biases terms in the DenseVariational layer to reduce the model complexity, which was sufficient for modeling and safe control as shown in the experimental results. First, we trained a BNN model and then used the posteriors of the weights in this BNN as the priors of the global model weights, to enhance the learning efficiency. Specifically, similar to the transfer learning approach for BNN training \cite{bao2020cdc}, we first trained an ANN model which shares the same architecture with the BNN model and then transferred the weights of the ANN model to the BNN model to improve the training efficiency of the BNN model. For model optimization, we used the Adam optimizer. 
Both the ANN and BNN model were trained for $30,000$ epochs with a batch size of $32$ and the initial learning rate of $1e-3$. The meta-learning of the BNN by Algorithm \ref{alg:meta} was executed for $100$ epochs with the initial learning rate of $1e-5$.

\begin{figure}[!htp]
    \centering
    \begin{subfigure}{0.49\columnwidth}
    \centering
    \includegraphics[width=\columnwidth]{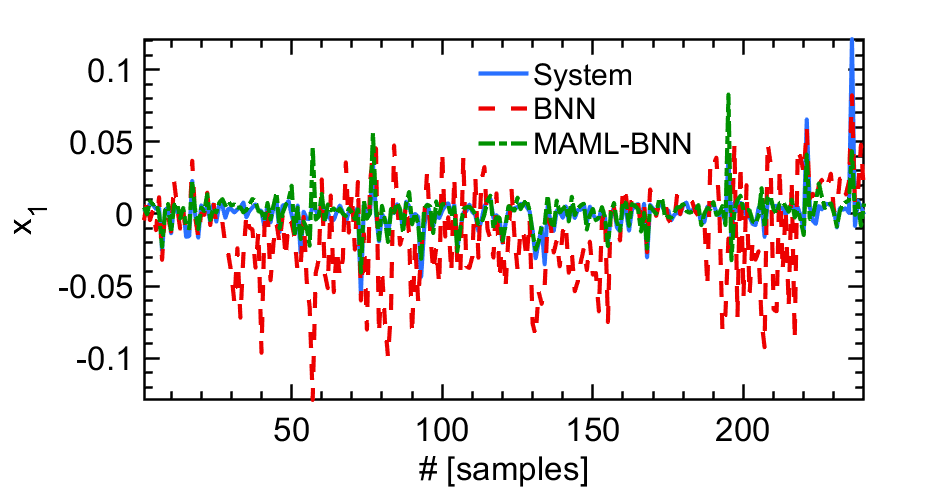}
    \caption{Comparison between the average models w.r.t. $x_{1}$.}
    \end{subfigure}
    \begin{subfigure}{0.49\columnwidth}
    \centering
    \includegraphics[width=\columnwidth]{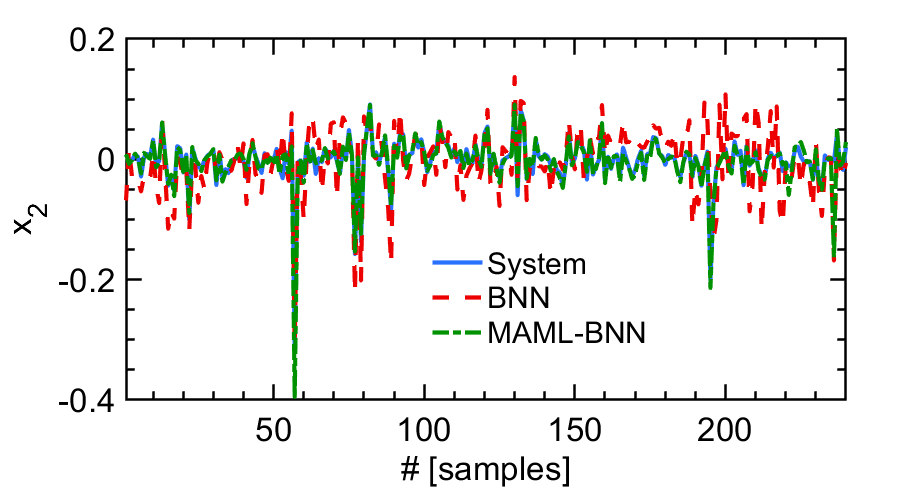}
    \caption{Comparison between the average models w.r.t. $x_{2}$.}
    \end{subfigure}
    \begin{subfigure}{0.49\columnwidth}
    \centering
    \includegraphics[width=\columnwidth]{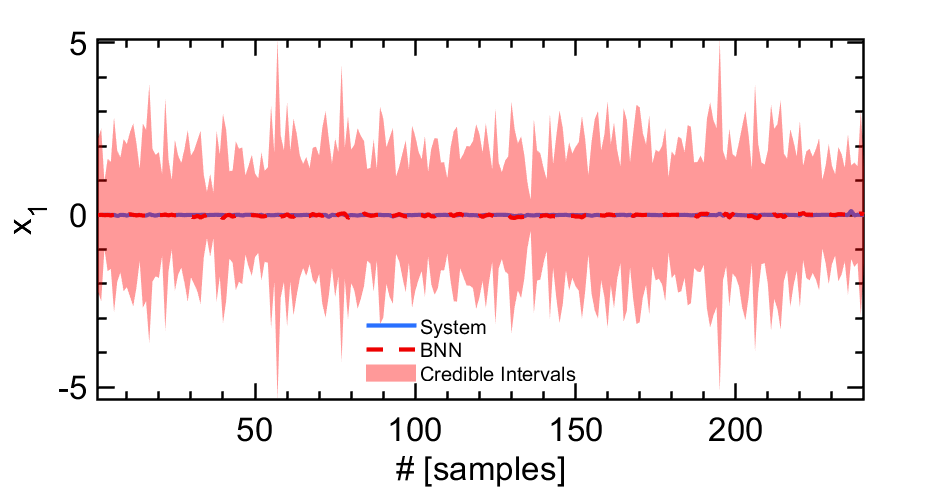}
    \caption{CIs by BNN w.r.t. $x_{1}$.}
    \end{subfigure}
        \begin{subfigure}{0.49\columnwidth}
    \centering
    \includegraphics[width=\columnwidth]{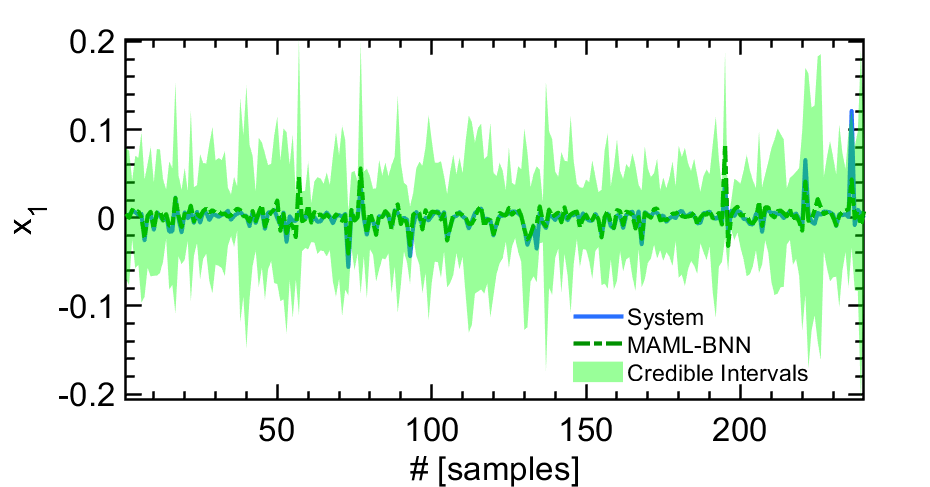}
    \caption{CIs of $x_{1}$ by MAML-BNN.}
    \end{subfigure}
    \begin{subfigure}{0.49\columnwidth}
    \centering
    \includegraphics[width=\columnwidth]{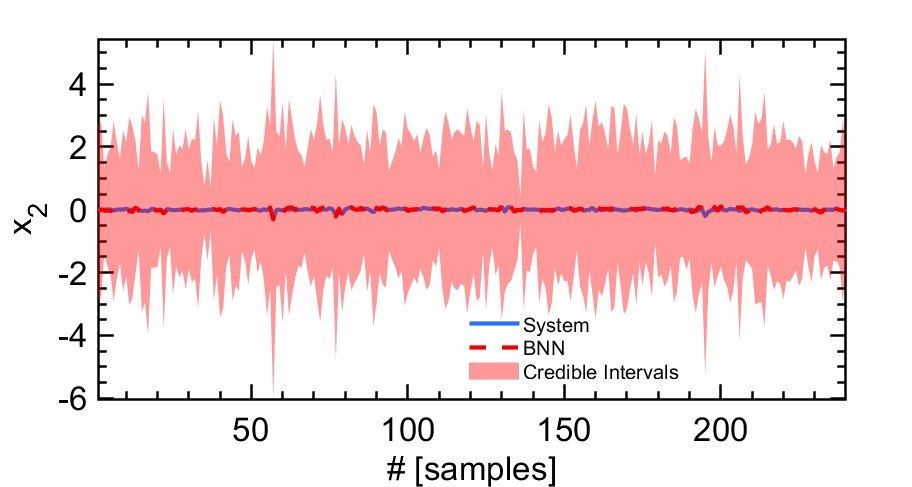}
    \caption{CIs by BNN w.r.t. $x_{2}$.}
    \end{subfigure}
    \begin{subfigure}{0.49\columnwidth}
    \centering
    \includegraphics[width=\columnwidth]{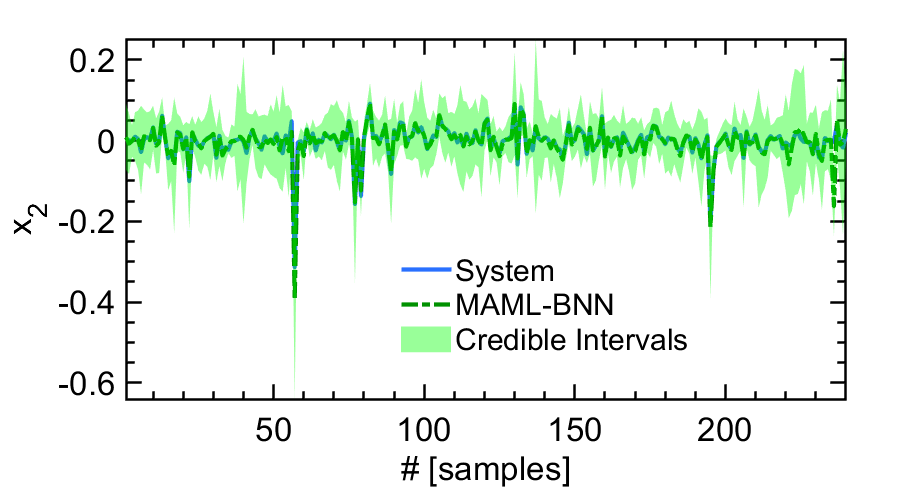}
    \caption{CIs of $x_{2}$ by MAML-BNN.}
    \end{subfigure}
    \caption{Comparison between BNN and MAML-BNN for the second example.}
    \label{fig:comp_num}
    \vspace{-2mm}
\end{figure}

\textbf{Results and discussion:}
The performance comparison between BNN and MAML of BNN on the testing set is shown in Fig. \ref{fig:comp_num}. It is noted that the accuracy of the average model by MAML-BNN is much higher than that by BNN and the credible intervals of MAML-BNN are far less conservative than those of BNN, which demonstrates that the proposed approach can improve the uncertainty quantification of BNN. Moreover, the closed-loop simulation results will demonstrate that the improved model will also improve the control performance. 

\subsubsection{Closed-loop Simulations}
Next, we use the MAML-BNN for scenario generation and design of an adaptive sMPC scheme. Similar to the first example, the control objective here is to stabilize the system while satisfying the system constraints.  

\begin{figure}[!htp]
\centering
\begin{subfigure}{0.49\columnwidth}
\centering
\includegraphics[width=\columnwidth]{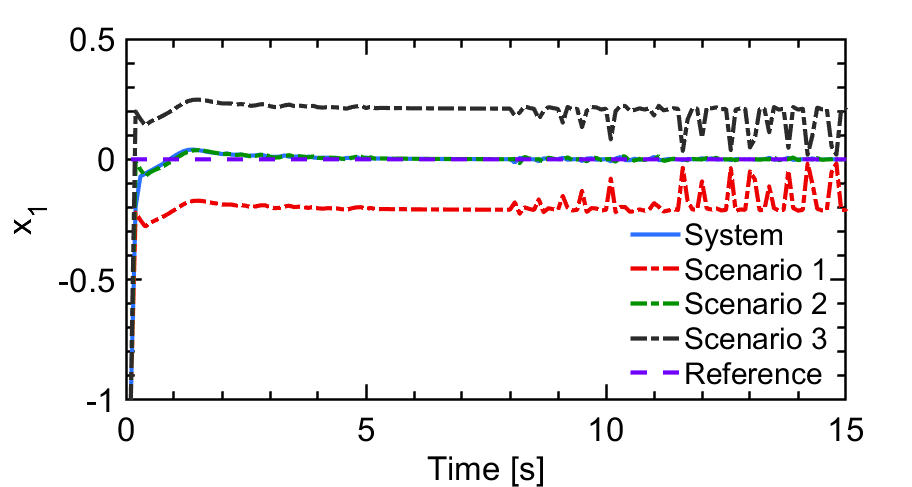}
\end{subfigure}
\begin{subfigure}{0.49\columnwidth}
\centering
\includegraphics[width=\columnwidth]{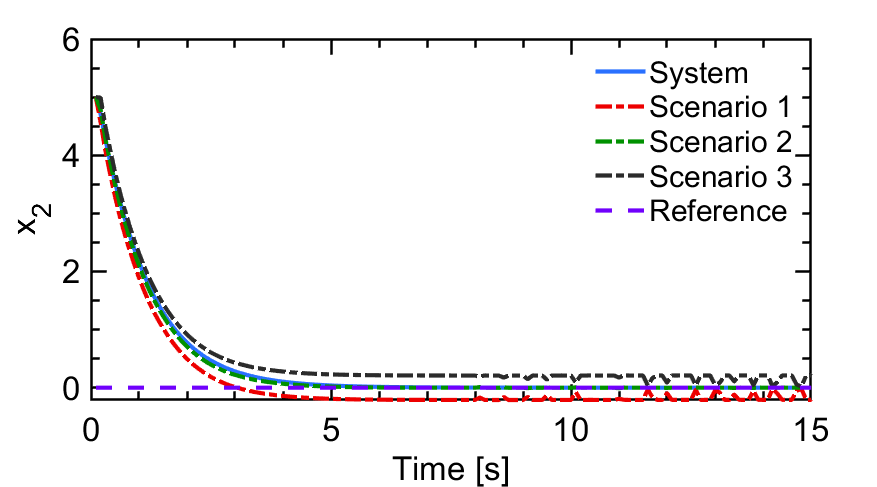}
\end{subfigure}
\begin{subfigure}{0.49\columnwidth}
\centering
\includegraphics[width=\columnwidth]{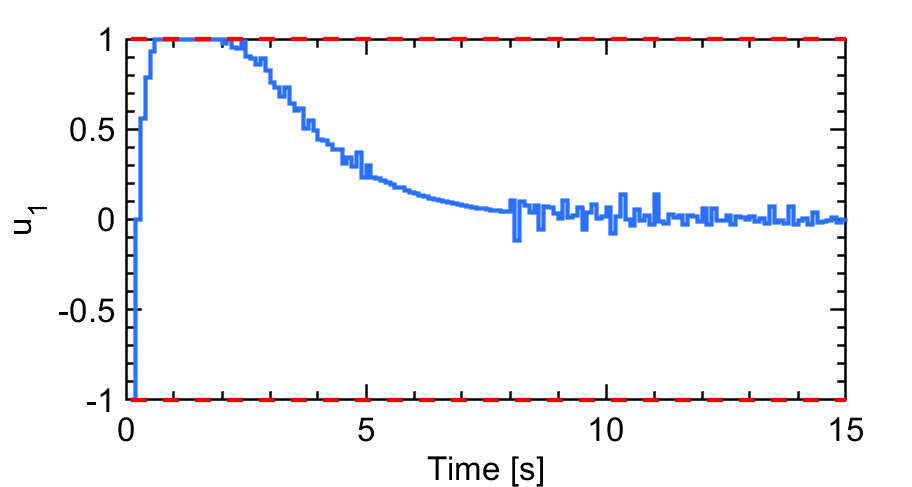}
\end{subfigure}
\begin{subfigure}{0.49\columnwidth}
\centering
\includegraphics[width=\columnwidth]{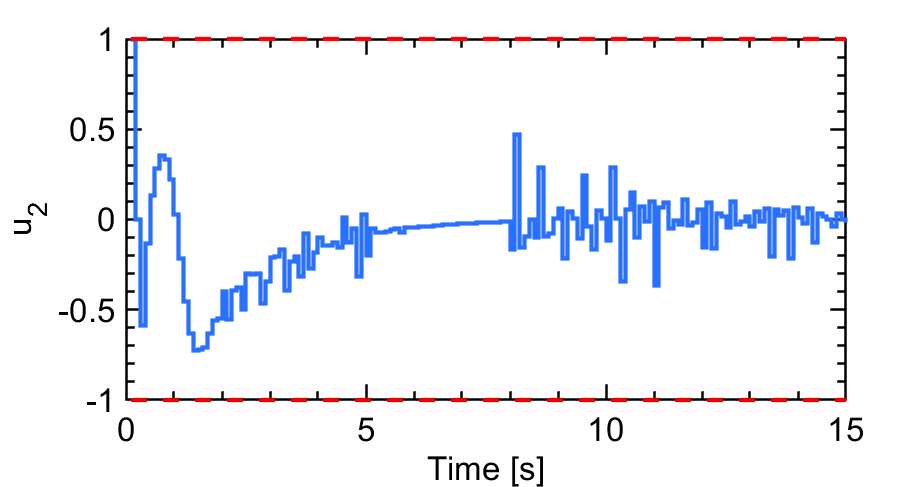}
\end{subfigure}
\caption{Closed-loop simulation results of the proposed sMPC with BNN by MAML for uncertainty estimation and weighted scenarios for system stabilization.} 
\label{fig:ctr_y_exp2}
\vspace{-4mm}
\end{figure}

At each time instant, we sampled $\bar{N}_{\text{MC}}=50$ models to estimate the mean $\mu_{g}$ and standard deviation $\sigma_{g}$. Subsequently, at each node of the scenario tree in the robust horizon, we used $\hat{\mu}_{g}$, $\hat{\mu}_{g}+3\hat{\sigma}_{g}$ and $\hat{\mu}_{g}-3\hat{\sigma}_{g}$ as the discrete scenarios branching from that node. Furthermore, the worst-case bounds of the mismatch are $|g_{1}|\leq 0.21$ and $|g_{2}|\leq 0.85$. When the predictions of the scenarios are out of the bounds of $g$, we use the bounds instead of the predictions and uniform distribution as the probability of scenarios. For sMPC design, we used a prediction horizon $N=7$ and a robust horizon $N_{r}=1$. The parameters of the adopted quadratic stage and terminal cost functions are $Q=I$, $R=100I$, and $P=I$, and the initial state is $x(0)=[-1;5]$. As shown in Fig. \ref{fig:ctr_y_exp2}, the proposed approach stabilized the system while satisfying the system constraints. Moreover, the trajectories of the three scenarios contain the trajectory of the system with a high probability, which guarantees safety.

\begin{figure}[!htp]
\centering
\begin{subfigure}{0.49\columnwidth}
\centering
\includegraphics[width=\columnwidth]{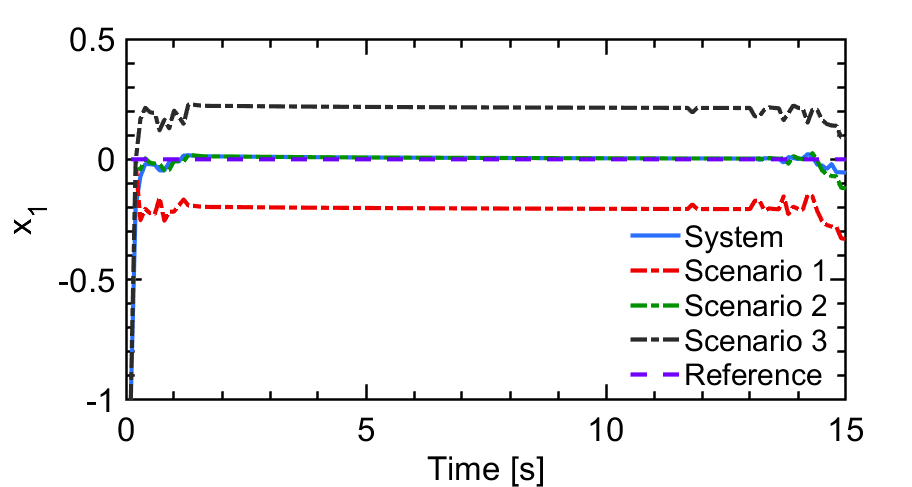}
\end{subfigure}
\begin{subfigure}{0.49\columnwidth}
\centering
\includegraphics[width=\columnwidth]{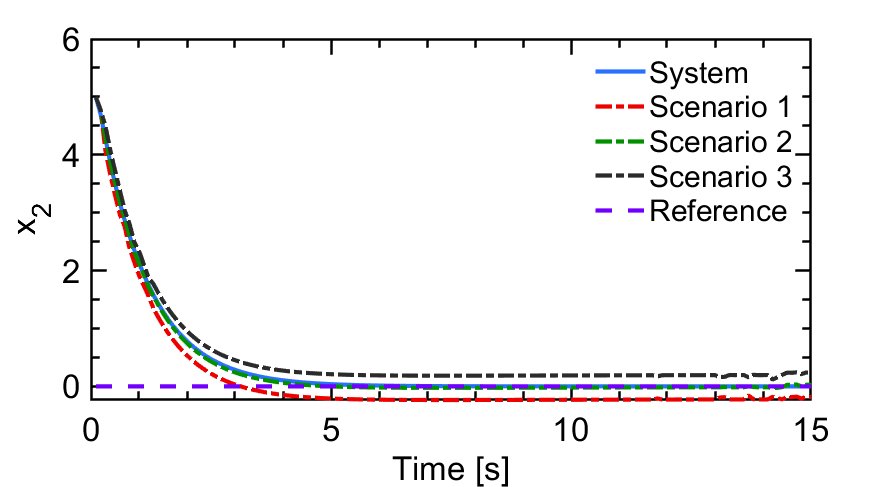}
\end{subfigure}
\caption{Closed-loop simulation results of the sMPC with BNN, clearly showing that the controller designed based on the BNN model failed to stabilize the system due to the inaccurate uncertainty quantification.} 
\label{fig:ctr_bnn}
\vspace{-2mm}
\end{figure} 

For comparison, we also examine the performance of the adaptive sMPC using the predictions of the plant-model mismatch by BNN without MAML. Fig. \ref{fig:ctr_bnn} illustrates that $x_{1}$ did not converge to $0$, which implies that the controller has failed to stabilize the system due to the inaccurate uncertainty quantification of the global BNN model.
\vspace{-2mm}
\section{Concluding Remarks}
\vspace{-2mm}
A model-agnostic meta-learning approach was presented to fine-tune BNN models online for sMPC design with safety guarantees. In particular, a global BNN model and an updating law for model adaptation were learned from data in the training phase. Then, the local BNN model adapted from the global model using the updating law was used to generate scenarios online for sMPC design purposes. To ensure safety, the behaviors of the generated scenarios contained the system behavior with high probability, and the constraints were enforced for all the scenarios. 
The closed-loop simulations demonstrated that the proposed approach improved model accuracy and control performance compared with sMPC designed based on a global BNN model. 

\vspace{-2mm}
\bibliography{ref}             
                                                   







\end{document}